\documentclass{aa}  
\usepackage{graphicx,amssymb,amsmath,hyperref,multirow}
\usepackage[usenames,dvipsnames]{xcolor}

\usepackage{rotating}
\usepackage{txfonts}
\usepackage{lscape}
\usepackage{amsmath}
\usepackage{natbib}
\usepackage{twoopt}
\usepackage{longtable}
\usepackage{multicol}
\usepackage{color}
\bibpunct{(}{)}{;}{a}{}{,} %% natbib format like A&A and ApJ
\newcommandtwoopt{\citeyearads}[3][][]%
{\href{http://adsabs.harvard.edu/abs/#3}{\citeyear[#1][#2]{#3}}}
\hypersetup{colorlinks,linkcolor=red,citecolor=blue,urlcolor=blue}

\begin{document}

\title{{Stellar twins {determine} the distance of the Pleiades}
\thanks{
Based on data products from observations made with ESO Telescopes at the La Silla Paranal Observatory under programme ID 096.D-0402(A).
Table 3 is only available in electronic form at the CDS via anonymous ftp to cdsarc.u-strasbg.fr (130.79.128.5) or via http://cdsweb.u-strasbg.fr/cgi-bin/qcat?J/A+A/}}

\author{
	 Thomas M\"adler\inst{\ref{ioa}},
	 Paula Jofr\'e\inst{\ref{ioa},\ref{port}}, 
	 Gerard Gilmore\inst{\ref{ioa}}, 
	 C. Clare Worley\inst{\ref{ioa}},  
	 Caroline Soubiran\inst{\ref{bord}}, \\
 Sergi Blanco-Cuaresma\inst{\ref{genf}},
  Keith Hawkins\inst{\ref{ioa}}, 
  Andrew R. Casey\inst{\ref{ioa}} 
		}

\authorrunning{T.~M\"adler et al. }
\titlerunning{Twins of the Pleiades}
\offprints{ \\ 
T. M\"adler, \email{tm513@cam.ac.uk}
P. Jofr\'e, \email{pjofre@ast.cam.ac.uk}
}

\institute{
	Institute of Astronomy, University of Cambridge, Madingley Road, Cambridge CB3 0HA, United Kingdom \label{ioa}
	\and 
	N\'ucleo de Astronom\'ia, Facultad de Ingenier\'ia, Universidad Diego Portales, 
         Av. Ejercito 441, Santiago, Chile \label{port}	
         \and
         Laboratoire d'Astrophysique de Bordeaux, Univ. Bordeaux, CNRS,UMR 5804, F-33615, Pessac, France\label{bord}
         \and
	 Observatoire de Gen\`eve, Universit\'e de Gen\`eve, CH-1290 Versoix, Switzerland \label{genf}         
}

   \date{}%Recived, acceptet

 \abstract {Since the release of the Hipparcos  catalog in 1997, the distance to the Pleiades open cluster has been heavily debated. 
The distance obtained from Hipparcos  and those by alternative methods differ by 10 to 15\%. 
As accurate stellar distances are key to understanding stellar structure and evolution, this dilemma puts the validity of stellar evolution models into question. 
Using our model--independent  method  to determine { parallaxes} based on twin stars, 
we report individual { parallaxes}  of 15 FGK type stars in the Pleiades in anticipation of the astrometric mission Gaia. 
{ These parallaxes give a mean cluster parallax of $7.42\pm0.09~\mathrm{mas}$ corresponding to a mean cluster distance of ${ 134.8\pm 1.7~\mathrm{pc}}$. }
This value agrees with the current results obtained from stellar evolution models. 
}


   \maketitle
%
%________________________________________________________________

\section{Introduction}

Open clusters like the Pleiades are of vital importance for our understanding of stellar evolution theory because of the assumption that stars form in groups within a common molecular cloud. 
 Since clusters are  composed of many stars, which are assumed to have the same chemical composition, distance and age, but  different initial masses and therefore residing at different evolutionary stages, they are ideal laboratories for testing stellar evolution models. 
An important output from these models is the intrinsic luminosity of a star. 
Given the theoretical luminosity and the observed flux, the model can be calibrated to  a star once the distance of the star is known. 
This is because  the observed flux of a star is proportional to the luminosity and inversely proportional  to the distance.

The Hipparcos mission \citep{ESA1997} measured to a high accuracy trigonometric parallaxes of about 120 000 stars in the solar neighbourhood. { In fact, one must be aware that the concept of  `stellar distance measurements' from Hipparcos or Gaia is incorrect, as the distance is not a true astrometrical observable. The measured quantity is the parallax, $\varpi$, which is a annual angular variation of an object with respect to distant stars and the parallax is related to the distance $d$  of that object by $d=1/\varpi$. }

\begin{figure*}
\begin{center}
\includegraphics[width=0.75\textwidth]{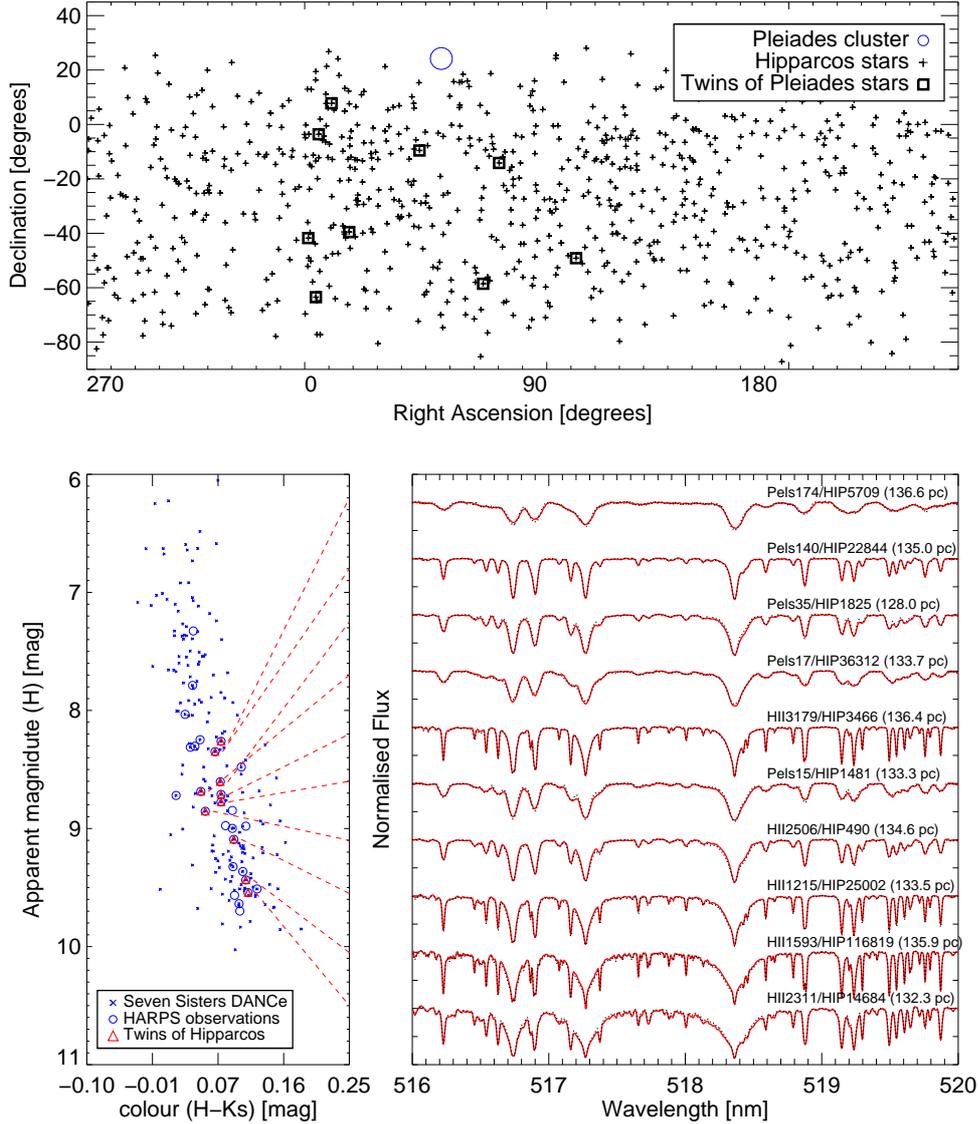}
\vspace{-1.5cm}
\caption{An illustration of the twin method applied to the Pleiades cluster. 
The coordinates of the stars used in this work are shown in the upper panel. 
The blue circle represents the location of the Pleiades cluster, the  black crosses are the reference Hipparcos field stars and the black squares are  the twins of the Pleiades stars found in the field.  
The bottom left panel shows the color-magnitude diagram of the Pleiades, including the stars of the Seven Sister DANCe catalogue \citep{2015A&A...577A.148B} for reference with crosses. 
Circles correspond to the stars for which we took HARPS spectra, and the red triangles correspond to  stars with  twins in the field. 
{ A selection  of Pleiades members with different spectra are shown  around the Mg I triplet} in red in the right bottom plot, together with a spectrum of a twin in the field as a dotted black line. 
The name of the Pleiades star and its corresponding  twin  are indicated, as well as its distance. 
 Fast rotators are  Pels 17, Pels 35 and Pels 174 which is seen in the greater broadening of their lines.  }
\label{fig:summary}
\end{center}
\end{figure*}

For most of the nearby open clusters, such as the Hyades,  the distances { obtained from Hipparcos parallaxes} agree well with those inferred from stellar evolution models. 
For the Pleiades, however, the Hipparcos {  parallax is  $8.32\pm~0.13~\mathrm{mas}$ corresponding to a distance  of  $ 120.3\pm1.5  ~\mathrm{pc}$}  \citep{floor2007, perrymanBook2008, floor2009, palmer2014} while other methods 
mainly  using theoretical modelling   estimate a distance of approximately $ 134~\mathrm{pc}$, { i.e. a parallax of about $7.46~\mathrm{mas}$.} 
These  methods include:  isochrone fitting \citep{Pinsonneault1998, Percival2005, An2007}, { empirical main sequence fitting \citep{StelloNissen2001}},  astrometric solutions that are alternative to Hipparcos  \citep{Makarov2002, Soderblom2005}, analysing  spectroscopic binaries \citep{Munari2004, Zwahlen2004, VAllGabaud2007,Groenewegen2007, Pan2004} and  long baseline interferometry \citep{Melis2014}. 
A recent  comparison between them can be found in  \cite{Melis2014} while an extensive discussion on the Pleiades distance controversy is in \cite{perrymanBook2008}.  

Recently, we proposed the twin method as a robust method for determining stellar distances \citep{2015MNRAS.453.1428J} that is independent of stellar modelling.
It assumes that if two stars at different locations in the sky have identical physical properties, then they are stellar twins. 
The difference in their apparent brightnesses  is  directly related to the difference in their distances. 
By knowing the distance of one star (e.g. from a Hipparcos parallax), it is then possible to know the distance of its twin.  
As the underlying assumption is that  twin stars are physically identical, therefore both stars must have the same spectra.
The twin method involves only observational quantities (the apparent brightness and the  observed spectra) in the distance determination. 
This is a great advantage  compared to non-astrometric distance determination methods relying on stellar evolution models and is  therefore a powerful independent technique for  estimating the distance of this important open cluster, provided spectra of { Pleiades member stars} are identical to spectra of field stars with accurate parallaxes. 

 A summary of this idea is illustrated in Fig.~\ref{fig:summary}, in which the spectra of some of the field stars  with accurate parallaxes located at different parts of the sky is identical to some of the spectra of Pleiades stars located at different parts of the HR diagram. One can measure the distance of each of the Pleiades spectroscopic twins independently.  In this paper we present in detail our procedure and discuss our results for the distance of the Pleiades.

\begin{table*}
\begin{center}
\begin{tabular}{||c|c|c|c|c|c|c|c|c|c|c||}\hline
star name 1& star name 2  & $\alpha_{2000}$ & $\delta_{2000}$ & $m_{V}$&$RV$ &$\boldsymbol{\varpi}$&$e_{\boldsymbol{\varpi}}$&SNR\\
 &   &  & & \tiny{$mag$}&\tiny{$[kms^{-1}]$} &\tiny{$[mas]$}&\tiny{$[mas]$}& \\\hline\hline
{$\dagger$  Pels 6}    &$\dagger$  HIP 16639        &{03 34 07.31}&{+24 20 40.0} &{9.61}&{1.03}&6.58&1.38&{29}\\\hline
{Pels 15}  &HIP 16979        &{03 38 22.57}&{+22 29 58.86}&{9.81}&{6.52}&6.08&1.82&{84}\\\hline
{Pels 17}  &HIP 17091        &{03 39 41.17}&{+23 17 27.1}&{9.93}&{5.44}&11.82&1.94&{94}\\\hline
{Pels 18}  &HIP 17044      &{03 39 13.47}&{+24 27 59.49}&{10.42}&{4.93}&10.19&2.19&{77}\\\hline
{$\dagger$  Pels 25}  &$\dagger$  HIP 17125        &{03 40 03.08}&{+27 44 25.83}&{9.56}&{-2.59}&9.19&1.66&{55}\\\hline\hline
{$\dagger$  Pels 26}  &$\dagger$  HIP 17481        &{03 44 44.85}&{+20 44 52.81}&{8.73}&{-3.88}&9.44&1.03&{71}\\\hline
{  Pels 27}  &$\dagger$  HIP 17289        &{03 42 04.72}&{+22 51 30.82}&{9.16}&{4.29}&7.65&1.50&{91}\\\hline
{Pels 35}  &HIP 17316        &{03 42 23.99}&{+21 28 24.57}&{9.85}&{7.34}&7.27&1.59&{121}\\\hline
{$\dagger$   Pels 42}  &$\dagger$  BD +25 610     &{03 45 45.12}&{+25 35 44.81}&{10.25}&{-1.07}&&&{80}\\\hline
{$\dagger$   Pels 70}  &$\dagger$  HIP 18154       &{03 52 53.47}&{+24 42 56.62}&{9.48}&{0.29}&10.13&1.66&{77}\\\hline\hline
{$\dagger$    Pels 86 }  &$\dagger$  HIP 18544       &{03 58 01.69}&{+20 40 36.48}&{9.37}&{10.34}&8.20&1.44&{90}\\\hline
{Pels 140}& HIP 17511       &{03 44 58.92}&{+22 01 56.82}&{9.43}&{5.80}&10.67&1.37&{101}\\\hline
{Pels 174}& HIP 18955     &{04 03 44.17}&{+22 56 39.40}&{9.67}&{6.71}&5.88&1.26&{61}\\\hline
{HII 430}   &                          &{03 44 43.98}&{+24 13 52.36}&{11.4}&{4.96}&&&{59}\\\hline
{$\dagger$   HII 948}   &$\dagger$  BD +22 549    &{03 46 12.69}&{+23 07 42.74}&{8.67}&{-6.50}&&&{37}\\\hline\hline
{HII 1215}&BD +23 527    &{03 46 53.75}&{+23 35 00.81}&{10.6}&{6.21}&&&{112}\\\hline
{HII 1593}&                          &{03 47 48.08}&{+23 13 05.11}&{11.2}&{7.23}&&&{65}\\\hline
{HII 1794}&BD +23 550    &{03 48 17.12}&{+23 53 25.4}&{10.2}&{5.63}&&&{64}\\\hline
{HII 1924}&   &{03 48 34.52}&{+23 26 05.3}&{10.7}&{5.83}&&&{96}\\\hline
{HII 2311}&                          &{03 49 28.74}&{+23 42 44.1}&{11.36}&{5.65}&&&{80}\\\hline\hline
{HII 2506}&BD +22 574    &{03 49 56.49}&{+23 13 07.03}&{10.2}&{6.03}&&&{80}\\\hline
{HII 3179}&BD +23 573    &{03 51 56.86}&{+23 54 07.08 }&{10.07}&{5.55}&&&{94}\\\hline
{}&$\dagger$  BD +20594    &{03 34 36.24 }&{+20 35 57.45  }&{10.85 }&{-20.35}&&&{}\\\hline
\end{tabular} 
\vspace{0.4cm}
\caption{ \label{tab:plei}Name, coordinates, apparent magnitudes in the visual, {\tt iSpec} radial velocity, parallax and { its standard error} according to \cite{floor2007} and SNR of the Pleiades stars analysed in this work.  Stars with $\dagger$   were rejected. }
\end{center}
\end{table*}

\section{Data} \label{s:selection}

The spectra used in this work were  taken by the HARPS instrument, which is fibre-fed by the Cassegrain focus of the 3.6m telescope in La Silla \citep{2003Msngr.114...20M}. 
The spectra were reduced by the HARPS Data Reduction Software (version 3.1).  
A combination of data taken from ESO public archive and observations taken  by ourselves during 4 nights November 2015  were considered in this analysis. 

The new observations contain spectra of 23 stars in the Pleiades field, which have spectral types FGK. 
This spectral type is important for enhancing the chances to find twins in the sample of field Hipparcos stars  available in the HARPS archive, which is composed mostly by FGK solar-type stars.  
Among these 23 stars, 11 have Hipparcos parallaxes. 
The apparent magnitudes of these stars vary from $V = 7$ to $11$~mag and in order to achieve a signal-to-noise (SNR) above 50 required for an accurate analysis, the exposure times of these stars were in the range of  1-2 hours. 
In several cases, shorter exposures of about 10 minutes were carried out before these long-time observations  to reject potential spectroscopic binaries showing double spectral lines. 
 All observations for each star were stacked using  {\tt iSpec} \citep{ispec2014}. 
 The basic information of the Pleiades targets can be found in Table~\ref{tab:plei}.

 The spectra of Pleiades stars were compared with a reference sample of  HARPS spectra of field stars that  have accurate  Hipparcos parallaxes.
 Most of the Hipparcos  field stars were taken from our previous work on twin distances \citep{2015MNRAS.453.1428J}, which were selected by cross-matching against the HARPS archive. 
 The cross-matched 664 stars were required to have  estimates of FGK spectral type \citep{2010A&A...515A.111S}.  
 In addition, we requested 62 stars  from the ESO archives\footnote{request number 192955} which matched the photometric properties of the Pleiades targets. 
Finally, 143 Hipparcos targets that were not observed so far with HARPS were added to the reference sample and observed during the same run as the Pleiades observations. 
They were selected to have similar photometric properties as the Pleiades targets, as a way to have more opportunities to find twins between both samples. 
 These stars are very bright and as such, exposures times of around 10 minutes were sufficient  to achieve high SNR. 
 The two additional requirements  of a non-saturated photometry in the $H$ and $K_s$ band and { a standard error of the Hipparcos parallax} smaller than 7\% reduced the reference sample to a total 598 stars. 
    The reference sample, which includes the name, coordinates, magnitude and parallax of the stars, as well as SNR, the date of HARPS observation and the programme ID in which the data were taken, can be found in Tab.~\ref{tab:field}.

The dataset was prepared for analysis using {\tt iSpec} functionalities resulting in  a homogeneous set of spectra. 
The spectra were normalised by fitting cubic splines to the pseudo-continuum, { corrected for radial velocity} by cross-correlation with the solar atlas of \cite{2000vnia.book.....H}, { cleaned from cosmic} and telluric lines,  sampled to common wavelength ranges and smoothed with a gaussian kernel to a lower resolution of 70,000. 
Although reducing the resolution is not imperative, the smooth spectra had fewer data points yet very high resolution allowing us to resolve the key spectral lines under study and perform a faster pixel-by-pixel comparison of the entire dataset.

\subsection{Selection of Pleiades members}
We assess the question on the cluster membership of the stars in the Pleiades field by comparing their radial velocities, which are listed in Table~\ref{tab:plei}.  
We considered the radial velocities determined with {\tt iSpec}, which was also used to process the spectra of the reference dataset.  
The Pleiades cluster has a radial velocity of approximately  $5.7~\mathrm{kms}^{-1}$ \citep{floor2009}. 
According to this value, the stars  Pels~6, Pels~25, Pels~26, Pels~42, Pels~70, Pels~86, HII~948 and BD~$+$20594 are not considered members of the cluster and were rejected from further analysis. 
After this membership classification we have in total 15 Pleiades members for the twin distance determination.

\subsection{Summary}
The twin method was applied to a sample of 613 FGK stars having high-resolution spectra and non-saturated 2MASS photometry \citep{2mass} in the near infrared  $K_s$ and $H$ bands.
The sample contained 15 Pleiades members and 598 fields stars distributed over the whole sky each with a Hipparcos parallax with an accuracy of better than 7\%.
The Pleiades spectra were  compared to the entire sample of field stars to search for potential stellar twins.

\section{Method}
In order to analyse homogeneously the slow and fast rotating stars in the Pleiades, we adopted a procedure to determine twin distances by comparing the spectra pixel-to-pixel rather than using equivalent widths as in \cite{2015MNRAS.453.1428J}. { The use of equivalent widths for high precision studies of solar twins for the determination of chemical abundances \citep{2006ApJ...641L.133M, 2014MNRAS.439.1028D, 2015A&A...579A..52N, 2016arXiv160604842S} is very common and has shown to be very powerful, but it reaches its limitation for  fast rotators ($v \sin i > 20~\mathrm{m/s}$). In our sample we have such fast rotators and therefore we developed a new procedure for this purpose. }
Key to this method is to take the difference of two spectra and evaluate the standard deviation $\sigma$ of these differences. 
 In this section we explain in detail our procedure.

\subsection{Twin distance determination}
\label{s:distance}
{ In the absence of interstellar extinction, there is the  well-known} relation between the apparent brightness, $m$ and fluxes, $F$, between two stars $1,2$:
\begin{equation}\label{flux_mag}
m_1 - m_2 = -2.5\log_{10}\left(\frac{F_1}{F_2}\right)
\end{equation}
As the flux is a function of the luminosity $L$ and the distance $d$ of a star, i.e. $F \propto L/d^2$, we have
\begin{equation}\label{withL}
m_1 - m_2 = 5\log_{10}\left(\frac{d_1}{d_2}\right) -2.5\log_{10}\left(\frac{L_1}{L_2}\right)
\end{equation}
The main assertion of the twin distance determination is that twins have the same intrinsic physical properties. 
Therefore twin stars must have the same intrinsic luminosity  and consequently the second term on the right hand side of Eq.~(\ref{withL}) vanishes. 
Another further consequence is that  twin stars must have the same intrinsic colour, allowing us to use the difference in observed colour as a proxy for interstellar extinction  on the observed magnitudes
\begin{equation}
E(B-V)_1 - E(B-V)_2 = R_V[(B-V)_1 - (B-V)_2] , 
\end{equation}
where $R_V$ is ratio of the of total-to-selective extinction in filter $V$, and $E(B-V)$ is the de-reddening.
Part of the Pleiades cluster is known to be obscured by the Merope Nebula, producing a reddening that is not constant to all stars \citep{1987ApJ...318..337S}.  Such errors due to extinction can be minimised by utilising apparent brightnesses in the $H$ and $K_s$ filters of 2MASS while having the further advantage of employing a fully homogeneous sample for the photometry.  
In these filters, we considered $R_{K_s} [(H-K_s)_1 - (H-K_s)_2] $  as a proxy for extinction, with  $R_{K_s} = 0.3$ according to \cite{2013MNRAS.430.2188Y}. 
Including this colour correction, twin distances can be determined from the following equation
\begin{equation}\label{tw_dist}
H_{1} - H_{2} - R_K[(H-K_s)_1 - (H-K_s)_2]  = 5\log_{10}\left(\frac{d_1}{d_2}\right) \;\;,
\end{equation}
Since for the reference stars we have direct measurements of the parallax $\boldsymbol{\varpi}=1/d$, we write Eq. \ref{tw_dist} as
\begin{equation}\label{tw_plx_0}
H_{1} - H_{2} - R_K[(H-K_s)_1 - (H-K_s)_2]  = 5\log_{10}\left(\frac{\boldsymbol{\varpi}_2}{\boldsymbol{\varpi}_1}\right) \;\;.
\end{equation} 
Note that the last equation only involves observable quantities. { 
Thus, our expression to determine twin parallaxes has the form 
 \begin{equation}\label{tw_plx}
\varpi_2 = \varpi_1 \times 10^{\frac{1}{5}\Bigg(H_{1} - H_{2} - R_K[(H-K_s)_1 - (H-K_s)_2] \bigg)} \,% = 5\log_{10}\left(\frac{\boldsymbol{\varpi}_2}{\boldsymbol{\varpi}_1}\right) \;\;.
\end{equation} 
where we determined the unknown parallax $\varpi_2$ given the knowledge of the reference parallax $\varpi_1$ and photometry in the $H$ and $K_s$ bands of the stars labelled with $1$ and $2$, respectively. }

\subsection{Spectral analysis: Assessment of spectral similarity}
\label{s:ilis}
We  compared each spectrum of the reference sample with each of the 15 Pleiades spectra  around  423 atomic transitions listed in \cite{2014A&A...564A.133J} for Fe and  in \cite{2015A&A...582A..81J} for Mg, Si, Ca, Ti, Sc, V, Cr Mn, Co and Ni, which have been used for the abundance determination of the Gaia benchmark stars \citep[see also][]{2015A&A...582A..49H, 2016arXiv160508229H}. Furthermore, the strong features of the wings of the Mg I b triplet and the Balmer lines H$\alpha $ and H$\beta$ were also taken into account.  These spectral regions are known to contain information on the stellar atmospheric parameters and therefore are well suited to assess the overall equality of  spectra of FGK stars.  

In theory, twin stars  have identical spectra. In practice, this does not happen because the spectra have noise and are observed under different conditions for different stars \citep{TGMET1, TGMET2}. 
Thus, critical to this work is a parameter, which we call $\sigma$, to  address the similarity of the spectra such that the twin { parallax} formula of Eq.~(\ref{tw_plx}) is satisfied within an accepted error in stellar distance. 
We define $\sigma$ as the standard deviation of the pixel-to-pixel difference between the spectra in the regions mentioned above. 
The smaller the value of $\sigma$, the more similar are  the two spectra.

\begin{figure}[!t]
\begin{center}
\includegraphics[scale=0.5]{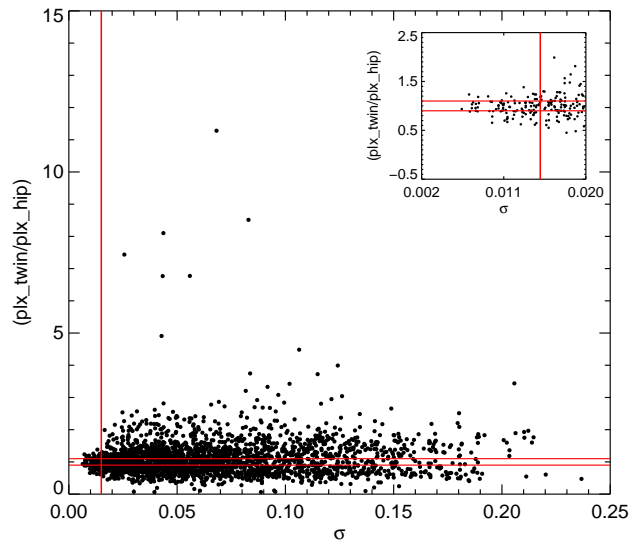}
\caption{\label{fig:sigma_devi} The behaviour of the ratio of twin parallax and Hipparcos parallax as a function of $\sigma$ using  the Hipparcos reference sample. The inlet shows a zoom of the region of the plot  at low $\sigma$.  The horizontal lines represent the range of 10\% difference in parallaxes while the vertical line the value of $\sigma = 0.015$.   }
\end{center}
\end{figure}

  \begin{figure*}[!t]
\begin{center}
\includegraphics[width=0.75\textwidth]{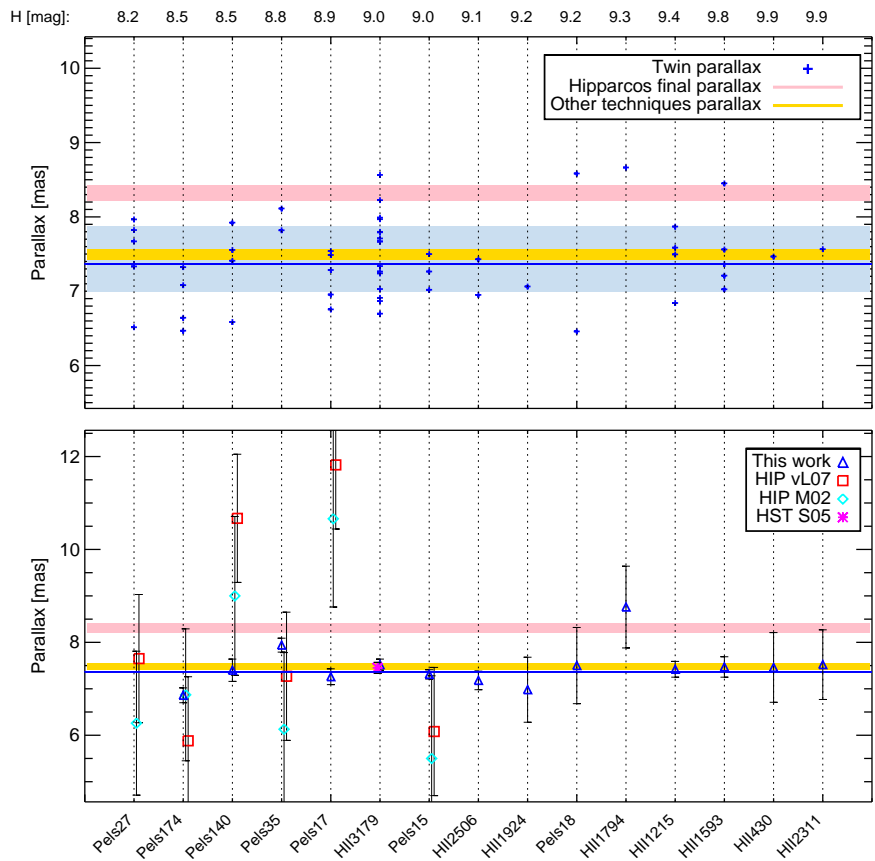}
\vspace{-2.cm}
\caption{
In both panels, the pink band represents the values of the Hipparcos solution within its margin of error while the yellow band represents the  range of the alternatives to Hipparcos.
The upper panels shows the { parallaxes of }  the 57 field twins  of each Pleiades member sorted by decreasing brightness, while  the median and dispersion of  the final twin { parallaxes}  are illustrated with the blue line and band, respectively.  
The lower panel displays the averaged individual twin { parallaxes}  (blue triangles), Hipparcos { parallaxes}  (red squares, HIP vL07), the alternative astrometric solution of \citep{Makarov2002} (cyan diamonds, HIP M02) and the parallax measurement with the Hubble Space Telescope of \citep{Soderblom2005} (pink star, S05). 
}\label{fig:dist_individual}
\end{center}
\end{figure*}

We investigated the threshold for $\sigma$ such that we obtain 10\% error { for the twin parallax.}
This threshold  was found by comparing the spectra with the twin distances of the entire dataset of reference field stars  with known Hipparcos parallaxes. 
That is, the ratio between the Hipparcos parallax $\boldsymbol{\varpi}_{\mathrm{HIP}}$ and the twin parallax $\boldsymbol{\varpi}_{\mathrm{twin}}$ was related to $\sigma$ for each pair of spectra in the data set. 
This can be seen  in Fig.~\ref{fig:sigma_devi}, where we plot the { ratio of the parallaxes obtained from Eq.~\ref{tw_dist} and  the Hipparcos parallax} as a function of $\sigma$ for the set of field stars. 
We found that when a pair had  very different spectra then  the value of $\sigma$ was  high and the value $(\boldsymbol{\varpi}_{\mathrm{HIP}}/\boldsymbol{\varpi}_{\mathrm{twin}})$ was also very different from unity. 
In the same way, if the spectra were very similar, $\sigma$ had a low value and the ratio between  Hipparcos and twin parallax was close to unity. 

Figure~\ref{fig:sigma_devi}  shows that for $\sigma < 0.015$, a typical error in { parallax} does not exceed 10\%. 
{ We refer to this parallax error $\Delta \varpi_\sigma$ as  the error of  the twinness, i.e. the relative error $\Delta \varpi_\sigma/\varpi<0.1$} for $\sigma=0.015$ (see below). 
This error is competitive to typical errors estimated using isochrone fitting when internal systematic errors on the models are neglected. 
The advantage here is that no stellar evolution models are required. 

Hence,  requiring $\sigma < 0.015$ for the Pleiades-field pairs yields individual { parallaxes} to the Pleiades targets with expected errors of 10\%.  
{ Note that this includes both the twin assumption error and the errors of the Hipparcos parallax.}

\begin{figure*}[!t]

\begin{center}
\includegraphics[width=\textwidth]{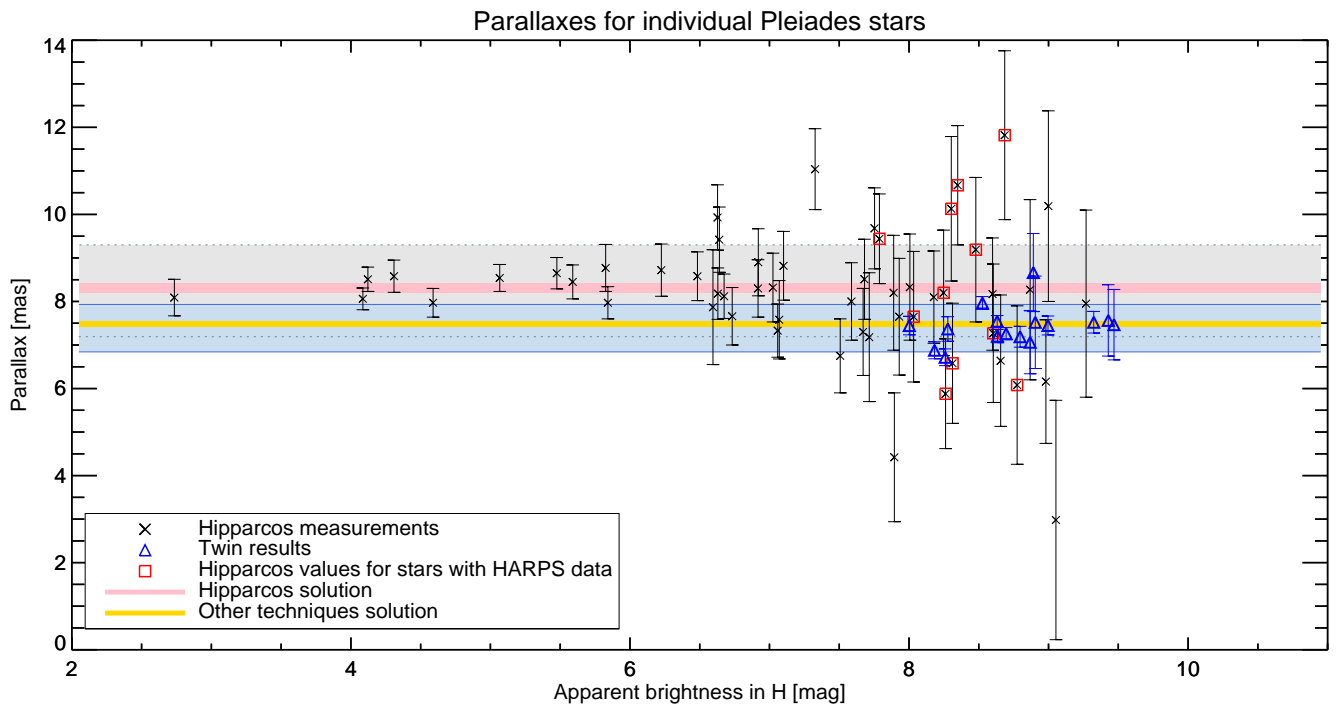}
\caption{ The { parallaxes}  of 54 Pleiades stars of the Hipparcos solution are  indicated with black crosses. 
The mean and standard deviation  of these Hipparcos { parallaxes}  is  shown by the grey band enclosed by  a dotted line. 
Eleven  of  the Hipparcos  stars have HARPS spectra and have been analysed in this work (red squares). 
 Our results  are indicated with blue triangles.  
 The standard deviation of these values are shown with the blue band enclosed by the blue line.  
 The pink band represents the values of the Hipparcos solution within its margin of error while the yellow band represents the  range of the alternatives to Hipparcos.   
}
\label{fig:hip}
\end{center}
\end{figure*}

\subsection{Uncertainties}

Below we summarize the sources of uncertainties involved in this work regarding our distance determinations.\\

\begin{enumerate}
{
\item {\it Reference parallax :} 
Our reference sample is comprised of only parallaxes with accuracies better than 7\% and typically of 3.5\%, therefore the traditional law of  error propagation can { be approximately applied} in our case. 
An uncertainty in the parallax associated to the reference star was propagated to the twin parallax  following \cite{2015MNRAS.453.1428J}. 
As discussed in that paper, typical  propagated errors are of the order of 5\%. \\
}
\item {\it Photometry :} In our sample, the median errors in the photometry of the stars for the $H$ and $K_s$  band are of 0.03 and 0.02~$\mathrm{mag}$, respectively.  
In \cite{2015MNRAS.453.1428J} we showed that this introduces a negligible error in {the final results of } less than 1\%. \\

\item {\it { twinness}:} The { error of the  twinness} $\Delta \varpi_\sigma/\varpi$ arises because the stars may not be exact twins and because the spectra have noise due to different observational condition. 
The latter was minimised by requiring high SNR of the observed spectra. 
As explained in the previous section the { error of the  twinness} is { less than} $10\%$ when choosing $\sigma<0.015$ in the comparison of the spectra. \\

\item {\it Multiple twins:}  If one star  has more than one twin, different values for its parallax can be estimated using each of these twins as reference. 
This gives a distribution of { almost independent parallax} measurements, with an associated mean with its standard error. 
This uncertainty is dependent on the number of twins found for a given star, where the larger the number of twins, the more accurate the distance.  
\end{enumerate}

In summary, if several twins are found for a target (which applies to the case of the Pleiades) the { reported  standard errors} correspond to the statistical error  (Point 4). 
If only one twin is found, then { the reported standard error} corresponds to the square root of the quadratic sum of the uncertainties explained in Points 1, 2 and 3.

\section{Results}

%We found 57 Pleiades-field-star pairs fulfilling this criterion.
Applying the cut of $\sigma<0.015$ to the difference of the spectra of the 8970 Pleiades-field-star-pairs\footnote{The number is the product of the number of 15 Pleiades stars and 598 field stars.} resulted in 57 twin pairs which are listed in Table~\ref{tab:twin_candidates}. 
These pairs were used to determine the individual { parallaxes} of the Pleiades members, which are indicated in Table \ref{tab:plx_dist_individual}.
Most of the Pleiades targets have more than one partner in the field. 
As  multiple twins to an individual Pleiades star must then share  the physical properties, these multiple partners should also be twins with each other.
We compared their Hipparcos and twin { parallax} and found an agreement of better than 10\%, which is consistent with the uncertainty obtained from the selection of $\sigma$.
This comparison between the Hipparcos field stars implies that the reference { parallax} of the field twins are accurate.  
It further confirms the analysis of  \cite{2016ApJS..222...19K} who found that Hipparcos { parallax} to FGK  field stars similar to Pleiades member stars  are in good agreement with stellar models.

\subsection{{ Parallax} of individual Pleiades stars}

Figure~\ref{fig:dist_individual}  shows the { parallax} of individual Pleiades stars, in which our results are compared with  the literature \citep{Makarov2002,Soderblom2005,floor2007}. 
The star HII~3179 has   been analysed by \cite{Soderblom2005} who performed parallax measurements using the Hubble Space Telescope. 
Our twin { parallax}  and the  parallax from \cite{Soderblom2005}  are in  excellent agreement at 2\%. 
This is encouraging given that these works are independent. 

%\red{The comparison for six of the Pleiades stars  having  Hipparcos measurements \citep{Makarov2002,floor2007} shows a significant  difference  for  all, but one star. 
%We note here that for the values of \cite{Makarov2002} and \cite{floor2007} we considered the inverse of their published parallax} and the uncertainties propagated according to Eq.~\ref{er_pro}. Although we are aware that this propagation of error is not accurate when the error in the parallax is large, the discussion here only gives a qualitative comparison of our results since our conclusions do not change if we instead compare parallaxes.

It is important to remark that not all the Hipparcos parallax measurements have the large error bars of the Hipparcos stars shown in Fig.~\ref{fig:dist_individual}. Indeed, several brighter {  Pleiades members} have significantly more accurate measurements. We show in Fig.~\ref{fig:hip} our Pleiades  twin { parallax} with Hipparcos  { parallaxes} of 54 Pleiades stars that are usually considered for the astrometric solution \citep[e.g.][]{Makarov2002, floor2007, palmer2014}. { The bright stars ($H<7~\mathrm{mag}$) have accurate Hipparcos parallaxes with a low star-to-star scatter of $0.5~\mathrm{mas}$ with an averaged parallax  of  $8.36\pm 0.09~\mathrm{mas}$}
(Fig.~\ref{fig:hip}), suggesting that these may be better candidates for comparison. 
However, this is currently not possible, since these stars are mainly of A and B spectral type,  to which the twin method  is much more difficult to apply. 
First,   the  HARPS public archive contains  mainly   FGK  stars spectra due to the bias towards searching for exoplanets around solar-like stars. 
Indeed, for the AMBRE project \citep{2014A&A...570A..68D}  70\% of the HARPS archive were found to be  FGK  stars while the remaining $30\%$  were either binaries, had too low signal-to-noise or were outside the FGK-parameter space. 
Second, $20\%$ of the brighter Pleiades stars are either variable or binary stars, so  their photometry could not be used to determine their distance with the twin method.  
Furthermore, AB  stars are more massive than FGK  stars and so have faster evolution implying that there are much fewer AB  stars in the sky compared to FGK stars. 
However,  the twin method can potentially be applied to any spectral-type, provided a good reference sample is available, as was recently shown for twin supernovae \citep{SNtwins}.

For fainter stars ($H  >7~\mathrm{mag}$) the situation in Fig.~\ref{fig:hip} is very different.  
{ The Hipparcos  parallaxes  have large star-to-star scatter of up to $1.8~\mathrm{mas}$ while the twin  parallax show a scatter of around $0.4~\mathrm{mas}$. 
For further comparison we calculate the weighted mean  parallax of the Hipparcos stars with $H>7~\mathrm{mag}$.
Taking the weighted mean is preferred because the parallax estimates of Hipparcos correlate with the errors of the measurements for fainter stars  \citep{floor2007,floor2009}.  
The weighted mean parallax obtained for the faint Hipparcos stars is  $8.07\pm 0.20~\mathrm{mas}$, which still 
agrees  within the errors  of  the value obtained from the  bright stars. 
In turn, calculating the weighed mean of the 15  twin parallaxes  gives a value of $7.42\pm0.09~\mathrm{mas}$.}

{ Since the literature regarding the Pleiades controversy is more commonly discussed in terms of distances in parsecs, we transform our results of the individual parallaxes and mean parallax of the cluster  to  distance using $d=1/\varpi$. Due to this non-linear relation it is clear that a parallax with a symmetric error results in a distance with an asymmetric error.  A recent discussion on how parallaxes, distances and their errors are related can be found in \citet{Bailer-Jones2015}, where the standard error propagation law of
\begin{equation}
\label{er_pro}
d\pm\Delta d = \frac{1}{\boldsymbol{\varpi}} \pm \frac{\Delta \boldsymbol{\varpi}}{\boldsymbol{\varpi}^2}\;\;,
\end{equation}
for translating errors in parallax to errors in distance can only be used when the  error of the parallax is less than 20\%. 
As both the derived twins parallaxes and the weighted mean parallax have relative errors well below 20\% we are can apply \eqref{er_pro}. 
The individual distances  of the Pleiades members and their errors are displayed in Table \ref{tab:plx_dist_individual}.  Our main result is thus that the weighed mean parallax of the 15 stars obtained with the twin method yields a  distance  of $ 134.8\pm 1.7~\mathrm{pc}$ to the Pleiades.}

\subsection{Lithium}
\begin{figure}[!t]
\begin{center}
\includegraphics[scale=0.6]{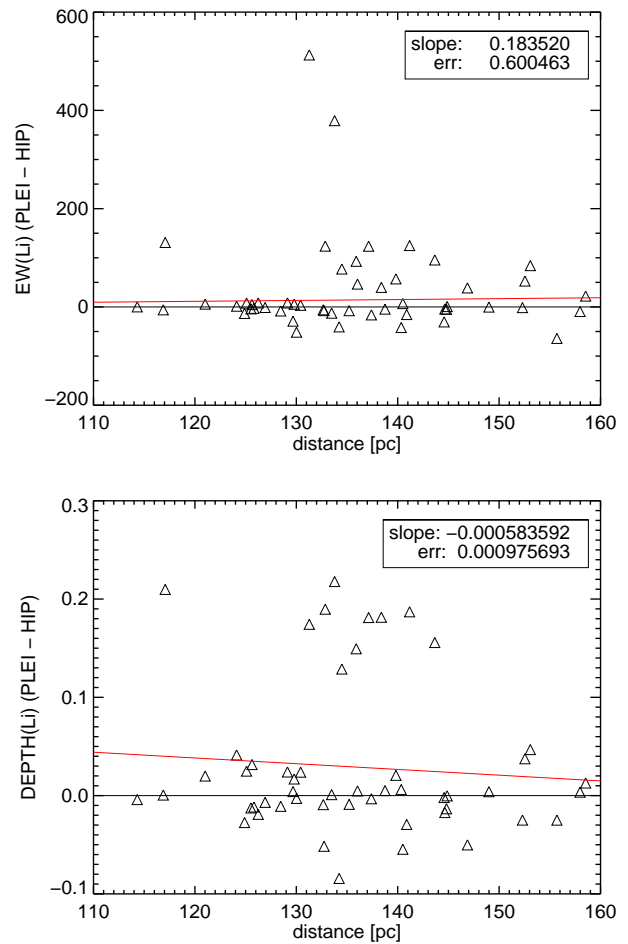}
\caption{\label{fig:lithium} The difference of the Li equivalent widths and depth  as a function of distance for the Pleiades-field twins. The red line indicated the linear fit of the data, with the slope and error of slope indicated in the legend. No correlation between difference of Li abundance and distance has been found.   }
\end{center}
\end{figure}

In our comparison of Pleiades and field stars the issue of age differences adding systematic uncertainties in the distance determination arise, because the cluster stars are all young while the field stars have different ages. In \cite{floor2009} it was shown { that the} colour-colour diagrams of the Pleiades and other young clusters were different with respect to the Hyades, suggesting an intrinsic difference in the luminosity and therefore distance modulus due to age. In order to study this potential systematic difference with our method, we looked at the Li $\lambda 6707\AA$ line in our spectra, taking into account that Li abundances can be used as an age-proxy \citep
{2014MNRAS.445.4306J, 2016ApJS..222...19K}, namely that a strong Li line indicates young stars. Using {\tt iSpec}, we measured the equivalent width and the depth of the Li $\lambda 6707\AA$ line in our sample of twins and correlated their difference with distance 
(see Fig.~\ref{fig:lithium}). A linear fit to the data was performed, showing no indication of systematic difference in distance for stars presenting same or different Li abundances \citep[see  also][for a  discussion on Li abundances of field stars with same photometric properties as  Pleiades member stars]{2016ApJS..222...19K}. Thus both young and older twins give the same result, showing that differential age effects are not significant in the method.

\begin{table*}
\begin{center}
\begin{tabular}{||l|l||}\hline
star name 1& Hipparcos name of twin candidate in the field  \\
Pleiades member &   \\\hline\hline

     Pels15 &   HIP43299,     HIP1481,     HIP3924   \\\hline

     Pels17 &   HIP43299,    HIP36312,     HIP2751,     HIP3924,     HIP2724   \\\hline

    Pels18 &    HIP6572,    HIP95149  \\\hline

   Pels35 &    HIP1481,     HIP1825   \\\hline

Pels 27 &HIP5099,    HIP41282,    HIP17838,    HIP18658,    HIP29932,    HIP19877 \\ \hline

   Pels140 &   HIP22844,     HIP1427,    HIP37844,    HIP46934   \\\hline

   Pels174 &   HIP41282,     HIP5709,     HIP7443,    HIP20350   \\\hline

  HII430 &   HIP14684  \\\hline

   HII1215 &    HIP3203,    HIP41587,   HIP109110,    HIP25002   \\\hline

  HII1593 &  HIP116819,    HIP48141,    HIP38041,    HIP91700,   HIP107805  \\\hline

   HII1794 &   HIP95149  \\\hline
   
   HII1924 &     HIP490   \\\hline
 
    HII2311 &   HIP14684   \\\hline

   HII2506 &     HIP490,     HIP1825   \\\hline

   HII3179 &   HIP33212,   HIP112117,    HIP72134,    HIP45685,    HIP38765,   \\   
                  &    HIP413,     HIP5280,    HIP26722,     HIP4747,     HIP5806,   \\
                  &   HIP23128,   HIP108859,   HIP115803,   HIP116106,     HIP3466,    HIP53094   \\\hline

\end{tabular} 
\vspace{0.5cm}
\caption{\label{tab:twin_candidates}Pleiades member stars and their twins in the field.  }
\end{center}

\end{table*}

\begin{table*}
\begin{center}
\begin{tabular}{| c | c | c  c  c  c  c |}
\hline
star name & Hipparcos  & $\boldsymbol{\varpi}$ & $e_{\boldsymbol{\varpi}}$ & $d$ & ${e}_d$ & \# \\
 &name &[mas]  &[mas] &[pc] &[pc]  & twins \\
 \hline
   Pels15 & HIP 16979 &  7.31 & 0.10   & 136.80 &  $^{+1.89}_{-1.94}$ &   3 \\\hline
   Pels17 &HIP 17091 &  7.26 & 0.17   & 138.13 &  $^{+3.08}_{-3.22}$ &   5 \\\hline
   Pels18 & HIP 17044&  7.50 & 0.82   & 136.69 &  $^{+13.15}_{-16.39}$ &   2 \\ \hline
   Pels 27 & HIP 17289 &  7.53 & 0.19   & 133.46 &  $^{+3.28}_{-3.45}$ &   6\\ \hline
   Pels35 &HIP 17316 &  7.94 & 0.15   & 126.00 &  $^{+2.36}_{-2.45}$ &   2 \\ \hline
  Pels140 &HIP 17511 &  7.40 & 0.24   & 135.75 &  $^{+4.22}_{-4.51}$ &   4 \\ \hline
  Pels174 & HIP 18955&  6.86 & 0.16   & 146.10 &  $^{+3.41}_{-3.57}$ &   4 \\ \hline
   HII430 & &  7.46 & 0.84   & 134.06 &  $^{+13.53}_{-16.95}$ &   1 \\\hline
   HII1215 & &  7.42 & 0.17   & 135.09 &  $^{+2.96}_{-3.09}$ &   4 \\ \hline
  HII1593 & &  7.47 & 0.22   & 134.48 &  $^{+3.83}_{-4.07}$ &   5 \\ \hline
 HII1794 & &  8.75 & 0.98   & 114.23 &  $^{+11.53}_{-14.44}$ &   1 \\ \hline
  HII1924 & &  6.98 & 0.78   & 143.23 &  $^{+14.45}_{-18.11}$ &   1 \\ \hline
  HII2311 & &  7.52 & 0.84   & 133.04 &  $^{+13.43}_{-16.82}$ &   1 \\ \hline  
  HII2506 & &  7.18 & 0.20   & 139.53 &  $^{+3.76}_{-3.98}$ &   2 \\ \hline
  HII3179 & &  7.52 & 0.12   & 133.47 &  $^{+2.09}_{-2.15}$ &  16 \\ \hline
\end{tabular} 
\caption{\label{tab:plx_dist_individual}  Our determinations of parallaxes and distances together with their error of the individual Pleiades members with twin candidates in the field. The Hipparcos name is given if known. }
\end{center}
\end{table*}

\section{Conclusions}

In this paper we have applied the twin method to determine distances of the Pleiades in a model-independent way. 
Our result of  $ 134.8\pm 1.7~\mathrm{pc}$, based on Hipparcos parallaxes of field stars,  disagrees with the value directly derived from Hipparcos parallaxes of Pleiades members \citep{floor2007, perrymanBook2008, floor2009, palmer2014}
but it is in good agreement with the value provided by the model-based methods mentioned earlier  \citep[]{Pinsonneault1998, Percival2005, VAllGabaud2007,Groenewegen2007,Pan2004, Melis2014}. 
Since most of these methods \citep[except][]{StelloNissen2001,Makarov2002, Soderblom2005,Melis2014} use stellar evolution models in their distance determination and because our observed stars are FGK  stars, our derived value implies that the current stellar evolution models for FGK  stars in the Pleiades are accurate.

Very soon the parallaxes from Gaia will confirm whether our prediction for the individual distances of the 15  Pleiades stars is correct, showing the power of the twin method in complementing Gaia and calibrating  distance scales. 
The Gaia Data Release 1 will { probably} also  show { if } the star-to-star dispersion for the fainter Pleiades targets remains. 
It will further confirm whether the Pleiades distance controversy is  a matter of individual distance accuracies,  cluster morphology,  astrometry,  or  the  approximations in stellar evolution models. 
In this study,  a  distance estimate to some of the most difficult Hipparcos targets within the Pleiades has been determined. 
At the end of the  Gaia mission,  accurate parallaxes will be available for almost every star for which there is  a high resolution spectrum. 
Nevertheless,  new larger telescopes are being constructed, such as the 40-meter E-ELT,  will give us high resolution spectra of very faint stars at the outskirts of our Galaxy and beyond.   
These ground-based instruments will provide new opportunities for complementing Gaia's astrometric solutions and so to continue to  climb the cosmic distance ladder with stellar twins.

\begin{acknowledgements}  
{ We are grateful to the referee U. Bastian whose comments helped to significantly to improve the article.}
It is our pleasure to thank T. Masseron, Q. Kral and  A. Bonsor for fruitful discussions on the subject.  This work was partly supported by the European Union FP7 programme through ERC grant number 320360. C.~W. acknowledges Leverhulme Trust through grant RPG-2012-541.  
Based on data obtained from the ESO Science Archive Facility and on data. This research has made use of the SIMBAD and WEBDA database.. 
\end{acknowledgements}

\begin{appendix}
\section{Data of field stars}
The information of the Hipparcos field stars used in this work is contained in Tab.~\ref{tab:field}. The name and coordinates of the stars are given, as well as the 2MASS photometry with its quality flag, the Hipparcos parallax from \cite{floor2007}, the SNR of the spectra, the date in which the spectra were taken and the program ID. 

\begin{table*}
\begin{center}
\hspace{-1cm}
\begin{small}
\begin{tabular}{ c |  c  c  c  c  c  c c c c c }
\hline \hline
star &  RA [deg] & DEC [deg] & H [mag]& K [mag]& $\boldsymbol{\varpi}$ [mas] &e\_$\boldsymbol{\varpi}$[mas]& SNR &date-obs &Program ID &2MASS Flag\\
\hline
 HIP100233  &304.938589 & -25.228242 &    5.86     &5.76 &   24.85  &   0.65 &  240.30  &2005-07-25 &  072.C-0488(E)  & AAA\\
 HIP101345  &308.097964  &-9.8536410   &  3.91    & 4.00   & 40.98    & 0.33  & 352.30  &2011-09-16  & 183.C-0972(A)  & DDE\\
 HIP101346  &308.099775  &6.51751652    & 7.17    & 7.10   & 10.07    & 0.84  & 127.90 & 2004-09-17  & 072.C-0488(E)  & AAA\\
 HIP101785  &309.431044  &-22.442887   &  6.35   &  6.31    &19.27   &  0.67  & 144.25 & 2011-05-29   &087.C-0831(A)   &AAA\\
 HIP101806  &309.466397  &-60.634323   &  6.12 &    6.03   & 20.01  &   0.66  & 234.35 & 2006-05-28   &072.C-0488(E)&   AAA\\
 \hline \hline
\end{tabular} 
\end{small}
\caption{\label{tab:field}  Properties of reference field Hipparcos stars used in this sample. A full version of this table can be found in electronic format. }
\end{center}
\end{table*}

\end{appendix}

\bibliographystyle{aa}
\bibliography{refs_plei}

\end{document}